\documentclass[sigplan,twocolumn]{acmart}
\renewcommand\footnotetextcopyrightpermission[1]{}
\AtBeginDocument{%
  }

\usepackage{tcolorbox}
\newtcolorbox{insightbox}{
  colback=gray!10,
  colframe=black,
  boxrule=1pt,
  arc=3pt,
  left=8pt,
  right=8pt,
  top=3pt,
  bottom=3pt,
  boxsep=2pt,
  before skip=5pt,
  after skip=5pt
}
\usepackage{filecontents}
\usepackage{tikz}

\usepackage{makecell}
\usepackage{multirow}
\usepackage{booktabs}
\usepackage[table]{xcolor}
\usepackage{graphicx}
\usepackage[linesnumbered, ruled, noend]{algorithm2e}

\usepackage{siunitx}
\usepackage{mhchem}

\usepackage{seqsplit}

\newcommand{\code}[1]{\texttt{\seqsplit{#1}}}

\setcopyright{acmlicensed}
\copyrightyear{2018}
\acmYear{2018}
\acmDOI{XXXXXXX.XXXXXXX}
\acmConference[Conference acronym 'XX]{Make sure to enter the correct
  conference title from your rights confirmation email}{June 03--05,
  2018}{Woodstock, NY}
\acmISBN{978-1-4503-XXXX-X/2018/06}

\begin{document}

\title{Request-Adaptive Execution for Efficient Hybrid LLM Inference}

\author{Fahao Chen}
\email{chenfh@ieee.org}
\affiliation{%
  \institution{Shandong University}
  \country{China}
}

\author{Xuzhe Yan}
\affiliation{%
  \institution{Xi’an Jiaotong University}
  \country{China}}
\email{Zedekiah5571@outlook.com}

\author{Qi Zhang}
\affiliation{%
  \institution{Xi’an Jiaotong University}
  \country{China}}
\email{zq0204@stu.xjtu.edu.cn}

\author{Peng Li}
\affiliation{%
  \institution{Xi’an Jiaotong University}
  \country{China}}
\email{pengli@xjtu.edu.cn}

\author{Dongxiao Yu}
\affiliation{%
 \institution{Shandong University}
 \country{China}}
\email{dxyu@sdu.edu.cn}

\author{Song Guo}
\affiliation{%
 \institution{Hong Kong University of Science and Technology}
 \country{Hong Kong SAR}}
\email{songguo@cse.ust.hk}

\renewcommand{\shortauthors}{Trovato et al.}

\begin{abstract}
Hybrid large language models (LLMs) have emerged as a promising way to combine the modeling capability of the Attention mechanism with the efficiency of state space models (SSMs). However, most existing hybrid LLMs still inherit a static design philosophy: once the Attention-to-SSM layout is chosen, every request must follow the same hybrid execution. This one-layout-fits-all design is inefficient for real-world serving workloads, where different requests rely on contextual information in different ways.

This paper presents \textsc{Flex}, a serving system that co-designs the model architecture and runtime for request-adaptive hybrid LLM inference. Its model architecture pairs Attention and SSM layers within each block and uses prefill-oriented gating to select a request-specific layout that remains fixed throughout decoding. Its serving runtime preserves batching efficiency under divergent layer choices through selective batching and enables correct prefix reuse through hybrid-aware context caching. Experiments on six datasets show that \textsc{Flex} achieves a better quality-efficiency trade-off than existing static hybrid baselines. System-level evaluation further shows that \textsc{Flex} improves the serving throughput by up to $3.87\times$ over baselines while substantially reducing inference latency.

\end{abstract}



\keywords{Large language models, State space models, Efficient serving}


\settopmatter{printfolios=true}
\maketitle
\pagestyle{plain}

\section{Introduction}
Large language models (LLMs) have become the foundation of modern AI services, powering applications such as search~\cite{sayyid2026llm,xu2024search}, coding assistants~\cite{kazemitabaar2024codeaid}, document analysis~\cite{ito2024tender,zhang2024detecting}, and autonomous agents~\cite{lin2024parrot,tan2025towards}. Most deployed LLMs are based on the Transformer architecture, where the attention mechanism provides strong global-context modeling but incurs quadratic computation and growing KV-cache memory during inference~\cite{zhang2024q, ye2025flashinfer, zhu2025sampleattention}. To reduce this cost, state space models (SSMs), such as Mamba~\cite{gu2023mamba,dao2024transformers}, have been introduced as efficient sequence modeling primitives. SSMs summarize prior tokens into compact recurrent states and update them in linear time. However, this compact recurrent state may limit the ability to recall context-dependent information, making SSMs less effective on tasks that require precise long-range context modeling~\cite{zhan2026overcoming}.

\begin{figure}[t]
    \centering
    \includegraphics[width=\linewidth]{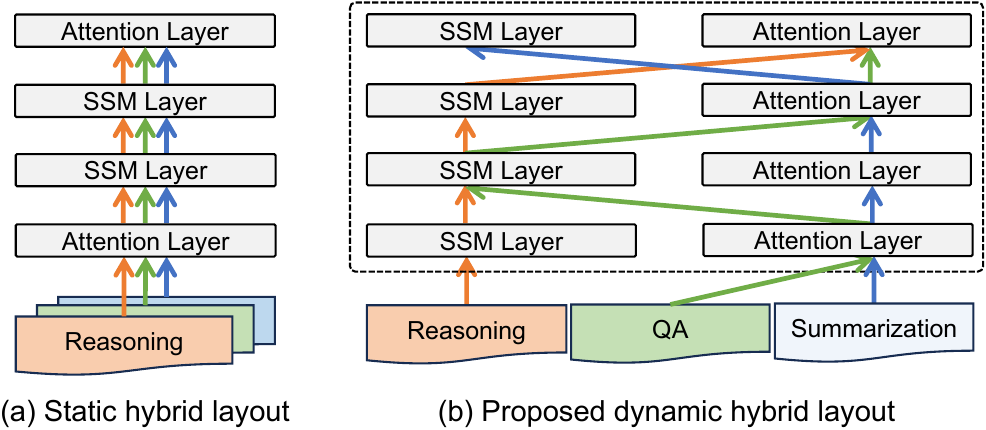}
    \caption{(a) Hybrid LLMs use the static Attention-to-SSM layout for all requests. (b) Request-adaptive hybrid layout allows different requests to activate different Attention and SSM execution. Colored lines indicate layer selections.}
    \label{fig:key_idea}
\end{figure}

To combine the complementary strengths of attention mechanisms and SSMs, recent works have proposed \emph{hybrid LLMs}\footnote{We use the term ``hybrid LLMs'' to broadly refer to LLMs that combine Attention layers with SSM layers within the same model.}, which interleave Attention layers and SSM layers within the same model to balance model quality and inference efficiency. Representative works such as Jamba~\cite{lenz2025jamba}, Mamba-LLaMA~\cite{wang2024mamba}, and Nemotron-H~\cite{blakeman2025nemotron} have demonstrated a better quality-efficiency trade-off than Attention-only or SSM-only architectures. A core design factor behind the performance of hybrid LLMs is the \textit{Attention-to-SSM layout}, which specifies not only how many Attention and SSM layers are used, but also where these layers are placed in the model.


Existing hybrid LLMs typically rely on hand-crafted architecture designs, where the Attention-to-SSM layout is selected based on empirical studies. In addition, this layout remains static for all requests once the model is constructed. For example, Jamba adopts an $a/m$ rotation strategy, inserting $a$ Attention layers for every $m$ SSM layers. As illustrated in \autoref{fig:key_idea}(a), such static designs assume requests with different characteristics, e.g., summarization, QA, and reasoning, benefit from the same hybrid structure. However, our preliminary experiments in \S\ref{sec:motivation} show that this assumption does not always hold in real-world inference workloads. A layout that achieves the best accuracy on a reasoning task such as ARC-Challenge can lead to suboptimal quality on a summarization task such as PubMed. This inconsistency arises because different workloads rely on contextual information in different ways. For example, reasoning tasks may only require selected Attention layers for option comparison, while long-document summarization often depends more heavily on fine-grained global context modeling to identify and preserve salient information. As a result, a static Attention-to-SSM layout can either overuse Attention for simple requests, wasting computation, or underuse Attention for context-sensitive requests, sacrificing output quality.

\textbf{Key Idea.} Our key idea is to make the hybrid LLM request-adaptive: rather than fixing the Attention-to-SSM layout, the model selects an appropriate hybrid layout for each incoming request. As illustrated in \autoref{fig:key_idea}(b), different requests can activate different Attention and SSM layers according to their context-modeling demands.
We realize this idea through algorithm-system co-design in \textsc{Flex}, an inference system that integrates a request-adaptive hybrid model with runtime optimizations for efficient serving. The hybrid model consists of multiple adaptive hybrid blocks, each containing a lightweight gate, paired Attention and SSM branches, and shared components such as the FFN. For each request, the gates select one branch per block, forming a request-specific execution layout that balances modeling capability and inference cost.

A key design choice is the gating granularity. A natural token-level gating design would allow each token to independently choose whether to execute the Attention or SSM layer in each hybrid block. However, this design is ill-suited for autoregressive decoding, where newly generated tokens rely on the cached results of previous tokens, i.e., Attention KVs and SSM states, to avoid recomputation. If tokens from the same request choose different layers within the same block, later tokens may not be able to directly reuse the Attention KVs or SSM states produced by earlier tokens, forcing the runtime to recompute previous tokens under the new layer choice.
\textsc{Flex} therefore adopts prefill-oriented gating mechanism (\S\ref{sec:gating}): during prefill, the gate uses all input tokens to make one request-level Attention and SSM choice for each hybrid block, and these choices are fixed throughout decoding. This allows the Attention KVs or SSM states built by previously generated tokens to be reused consistently.

At the system level, request-adaptive layouts introduce two challenges: requests within the same batch may select different branches, and requests sharing a textual prefix may require incompatible layer-wise caches. The serving runtime in \textsc{Flex} addresses these challenges through the following two designs to preserve batching efficiency and ensure correct context reuse.


\noindent\textbf{Selective Batching Engine.} To maintain high inference throughput, \textsc{Flex} must efficiently batch multiple requests for parallel execution. However, requests in the same batch may select different branches at each hybrid block, which breaks the traditional batching assumption of conventional serving engines, where all requests in a batch follow the same operator sequence~\cite{kwon2023efficient, sun2024llumnix, du2025prefillonly}. A naive solution would either serialize or execute block computations separately. The former loses parallelism, while the latter fragments the original batch and prevents request-independent operations, such as normalization and FFN computation, from being executed as a single compact batch~\cite{zhu2025nanoflow}.

\textsc{Flex} addresses this with selective batching engine (\S\ref{sec:batching}). The key idea is to keep branch-independent operations on the original packed batch, and split requests only for the branch-dependent Attention and SSM computation. At each block, it partitions requests into compact Attention and SSM sub-batches, executes them with their corresponding kernels, and then merges outputs back into the original batch layout. Since two branches may have different execution costs, \textsc{Flex} further applies load-aware branch coordination by assigning more resources to heavier branch, reducing synchronization stalls before merging.

\noindent\textbf{Hybrid-Aware Context Caching.} Real-world LLM serving workloads often contain repeated long prefixes, such as system prompts, document contexts, and multi-turn conversation histories. Recomputing these prefixes for every request wastes substantial prefill computation, especially in long-context scenarios. Context caching~\cite{jin2025ragcache, yao2025cacheblend, yu2025ic} addresses this issue by identifying matched textual prefixes and reusing their previously generated caches. Existing techniques typically assume that a matched textual prefix uniquely determines all reusable caches. However, this assumption no longer holds for \textsc{Flex}: even when two requests share the same prefix, they may select different Attention and SSM layers across hybrid blocks, producing different layer-wise caches. Moreover, conventional cache eviction operates at the context level, where evicting a prefix removes all layer-wise caches associated with it. This coarse-grained eviction is inefficient for request-adaptive hybrid execution, because only a subset of layer caches under a matched context may be frequently reused. Evicting the whole context can reduce future cache hit ratios, while retaining it may keep many rarely used layer caches in memory.

To address these limitations, \textsc{Flex} introduces the hybrid-aware context caching (\S\ref{sec:context}). Its core mechanism, a two-level radix tree structure, decouples textual-prefix matching from layer-wise cache matching. After identifying a shared textual prefix, it further determines which layer-wise caches are compatible with the current hybrid layout. Built on this structure, \textsc{Flex} further performs priority-based eviction at the layer-wise cache level, accounting for context sharing potential, layer-wise reuse likelihood, and recomputation cost. As a result, \textsc{Flex} preserves useful shared prefixes while evicting low-value layer caches, enabling correct and efficient context reuse for dynamic hybrid LLM serving.

\textbf{Results:} We evaluate \textsc{Flex} on the NVIDIA H100 GPU using six datasets spanning reasoning, QA, and summarization tasks. Experiments show that the proposed hybrid model achieves a better quality-efficiency trade-off than homogeneous and static hybrid baselines, obtaining the best results on 13 out of 18 quality metrics while reducing estimated TFLOPs/token by up to 67.2\% relative to LLaMA. At the system level, \textsc{Flex} improves up to $3.87\times$ the serving throughput of vLLM while substantially reducing time-to-first-token (TTFT) and time-per-output-token (TPOT).

\begin{figure}[t]
    \centering
    \includegraphics[width=\linewidth]{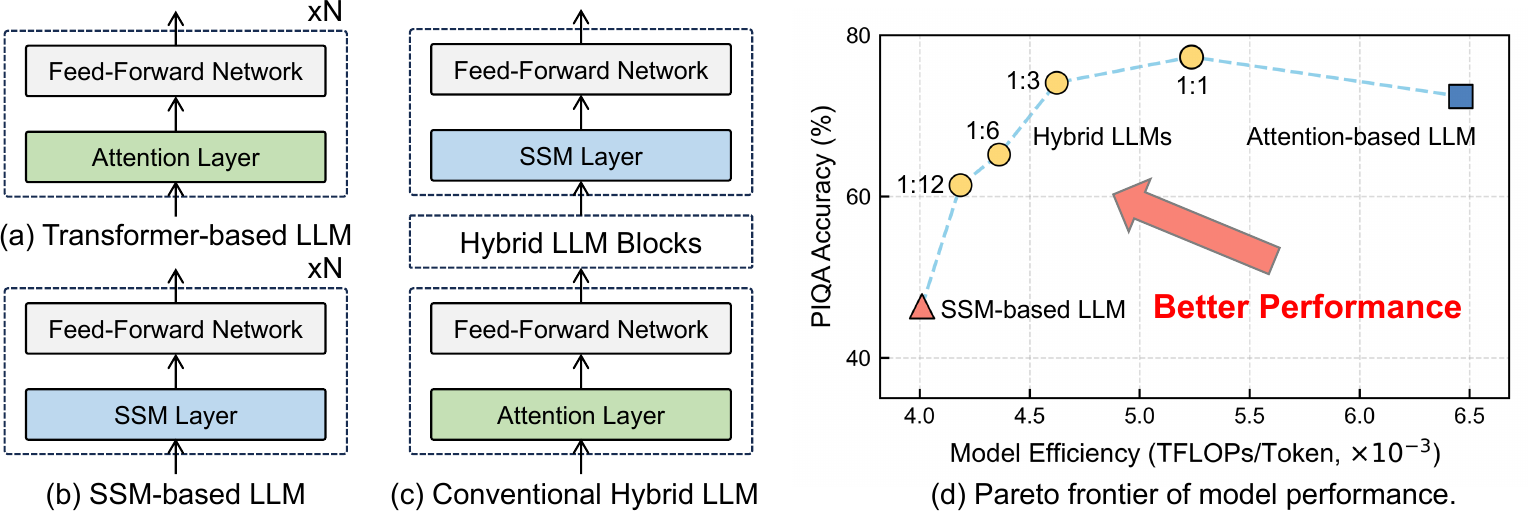}
    \caption{Overview and quality-efficiency trade-off of conventional hybrid LLMs. (a)-(c) Illustrations of Attention-based, SSM-based, and conventional hybrid architectures. (d) PIQA accuracy versus model efficiency under different Attention-to-SSM layouts. All models have 3B parameters.}
    \label{fig:background}
\end{figure}

\section{Background and Motivation}
\subsection{Background}
\noindent\textbf{Attention-based LLMs.} Attention-based LLMs rely on multi-head self-attention to capture global token relationships~\cite{vaswani2017attention}. Given the hidden representations $X$, the Attention layer projects them into queries ($Q$), keys ($K$), and values ($V$) tensors and computes:
\begin{align}
    \mathrm{Attn}(Q,K,V) = \mathrm{softmax}\left(\frac{QK^\top}{\sqrt{d_h}}\right)V,
\end{align}
where $d_{h}$ denotes the head dimension. This dense interaction allows each token to directly aggregate information from previous tokens, providing strong capability for reasoning and long-context understanding. However, the computation of an Attention layer grows quadratically with the prompt length during prefill~\cite{zhang2024q, ye2025flashinfer, zhu2025sampleattention, zou2025axcore, zhou2026sas}. During decoding, it needs to maintain KV caches for preceding tokens, introducing increasing memory and computation costs as the context grows.

\noindent\textbf{SSM-based LLMs.} State space model (SSM), such as Mamba model~\cite{gu2023mamba, dao2024transformers}, provide an efficient alternative by summarizing historical information into compact recurrent states:
\begin{align}
    h_t = \bar{A}_t h_{t-1} + \bar{B}_t x_t,\qquad
    y_t = C_t^\top h_t + D x_t,
\end{align}
where $h_t$ is the SSM state at time $t$, $\bar{A}_t\in \mathbb{R}^{(N,N)}$ and $\bar{B}_t\in \mathbb{R}^{(N,1)}$ define the input-dependent state update, $C_t\in \mathbb{R}^{(N,1)}$ maps the state to the output, and $D$ is a skip connection. Unlike conventional linear time-invariant SSMs, Mamba makes the state update input-dependent, enabling selective preservation of information from the input sequence~\cite{gu2023mamba}. Since an SSM layer compresses historical tokens into a bounded-size state, it avoids the growing Attention KV cache required by Attention-based LLMs and supports efficient long-sequence inference. Mamba-2 further improves the efficiency of selective SSM computation through state-space duality~\cite{dao2024transformers}. However, this compact recurrent state may limit the ability to recall context-dependent information, making SSMs less effective on tasks that require precise long-range context modeling~\cite{zhan2026overcoming}.

\noindent\textbf{Hybrid LLMs.} Hybrid LLMs combine Attention layers and SSM layers within the same model to balance modeling capability and inference efficiency, as shown in \autoref{fig:background}. A key design factor is the \emph{Attention-to-SSM layout}, which determines how frequently Attention and SSM layers are used. In \autoref{fig:background}(d), we evaluate layouts of $1{:}1$, $1{:}3$, $1{:}6$ and $1{:}12$, where $1{:}k$ layout denotes one Attention layer for every $k$ SSM layers. The SSM-only configuration achieves the lowest computation cost but suffers substantial accuracy degradation, whereas the Attention-only configuration incurs much higher cost. Several hybrid layouts achieve higher accuracy than both endpoints with moderate computation cost, demonstrating the advantage of combining Attention and SSM layers. Moreover, the performance varies noticeably across layouts, indicating that the hybrid layout is a critical factor in determining the quality-efficiency trade-off.

\subsection{Motivation}\label{sec:motivation}
The quality-efficiency trade-off of a hybrid LLM is largely determined by its \emph{Attention-to-SSM layouts}. Existing hybrid LLMs determine this configuration based on empirical studies and keep it static during model serving. For example, Jamba~\cite{lenz2025jamba} periodically inserts Attention layers into an SSM-dominated backbone, while Nemotron-H~\cite{blakeman2025nemotron} fixes a predefined distribution of Attention and SSM layers. Once constructed, all requests are forced to execute the same layout, regardless of their input characteristics or contextual requirements. We find that such static layouts are insufficient for diverse inference workloads: different tasks, and even different requests from the same task, can prefer different uses of Attention and SSM computation.

\begin{figure}[t]
    \centering
    \includegraphics[width=\linewidth]{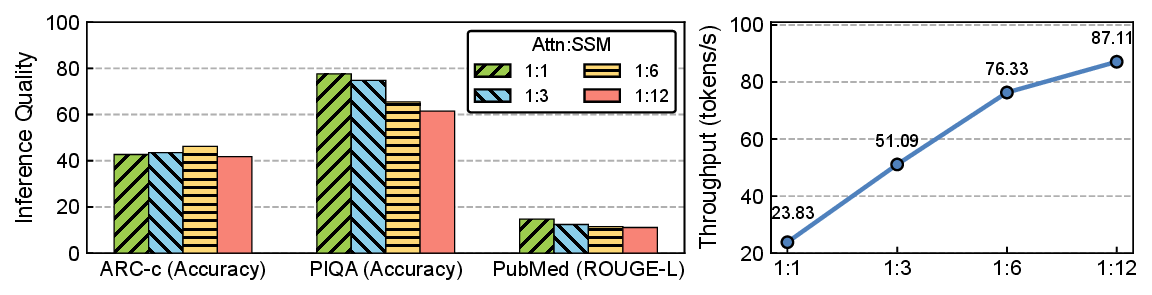}
    \caption{Quality and throughput of static hybrid LLMs with different Attention-to-SSM layouts.}
    \label{fig:motivation_1}
\end{figure}
\begin{insightbox}
\noindent\textbf{Observation 1: No single Attention-to-SSM layout can achieve the best performance across different workloads.} 
\end{insightbox}
We first study whether a single static hybrid layout can achieve a consistently good quality-efficiency trade-off across different workloads. We use a set of static hybrid models from~\cite{wang2025systematic}, which combine Attention layers with SSM layers under different Attention-to-SSM layouts ranging from $1{:}1$ to $1{:}12$. We evaluate these layouts on a single NVIDIA H100 GPU using three representative datasets: ARC-Challenge for reasoning, PIQA for commonsense QA, and PubMed for long-document summarization. We report task quality together with generation throughput.

As shown in \autoref{fig:motivation_1}, different workloads favor different hybrid layouts. On ARC-Challenge, the $1{:}6$ hybrid layout achieves the highest accuracy, suggesting that reasoning benefits from sufficient SSM efficiency while retaining selected Attention computation. On PIQA and PubMed, the $1{:}1$ hybrid layout performs best, reflecting its stronger demand for fine-grained long-context modeling. Meanwhile, increasing the proportion of SSM layers significantly improves throughput, from 23.83 tokens/s under $1{:}1$ to 87.11 tokens/s under $1{:}12$. These results show that the Attention-to-SSM layout directly affects both task quality and serving efficiency, but no single static layout is consistently optimal across diverse workloads.

\begin{figure}[t]
    \centering
    \includegraphics[width=\linewidth]{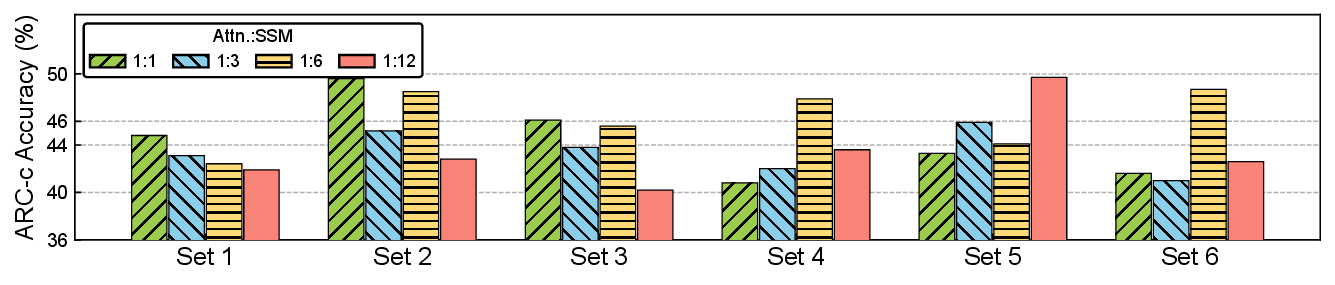}
    \caption{Accuracy of different static Attention-to-SSM layouts on six disjoint ARC-Challenge subsets.}
    \label{fig:motivation_2}
\end{figure}
\begin{insightbox}
    \noindent\textbf{Observation 2: Even within the same workload, different requests prefer different Attention-to-SSM layouts.}
\end{insightbox}
Even within the same task type, different requests can still prefer different hybrid structures. To show this, we further study ARC-Challenge by partitioning the evaluation set into six disjoint subsets, each containing 100 requests, and evaluate the same four static Attention-to-SSM layouts on each subset. As shown in \autoref{fig:motivation_2}, the best layout changes across subsets. For example, Set 1 and Set 3 favor more Attention-heavy settings, while Set 4 and Set 6 achieve the best results under the $1{:}6$ hybrid layout. Set 5 instead performs best with the $1{:}12$ layout. Even within a single reasoning benchmark, there is no universally optimal static hybrid structure. This result suggests that the preferred Attention-to-SSM layout depends not only on the task category, but also on the specific requests within the same dataset. Therefore, a static hybrid layout is fundamentally limited, motivating a dynamic hybrid architecture that can adapt its structure to each request.

\section{System Overview} \label{sec:overview}
\begin{figure}[t]
    \centering
    \includegraphics[width=\linewidth]{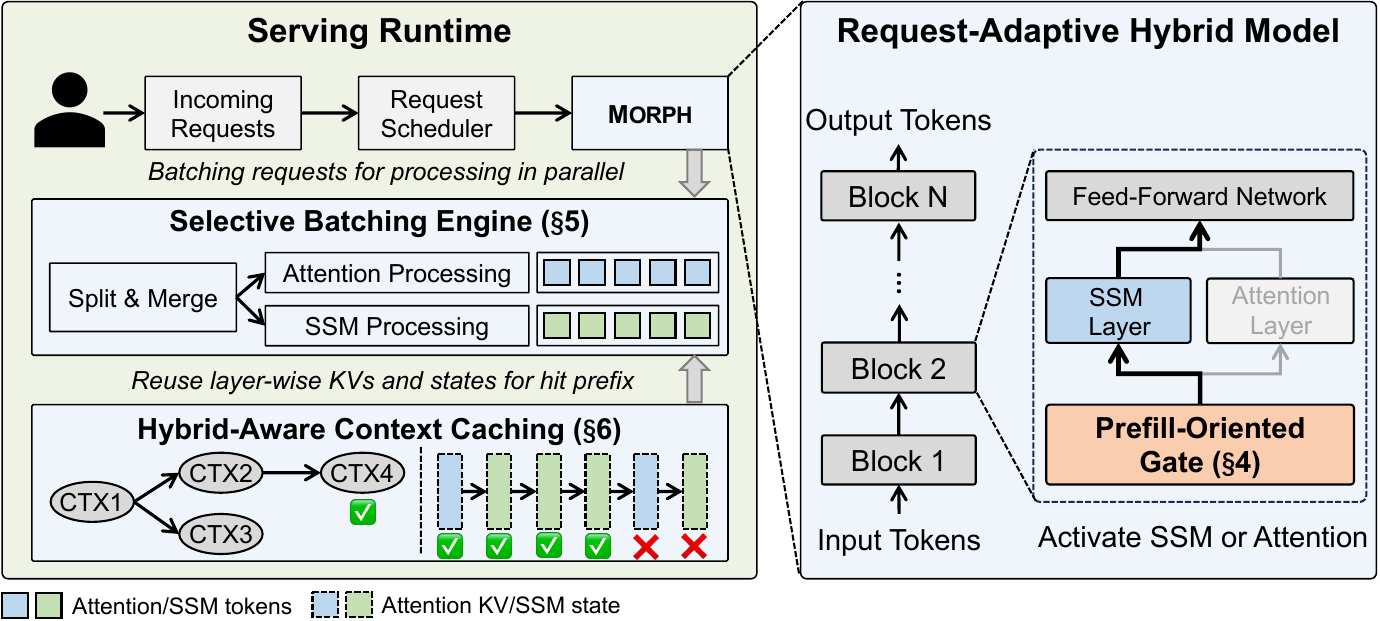}
    \caption{System overview of \textsc{Flex}. For clarity, standard components in the model, e.g., normalization, are omitted.}
    \label{fig:overview}
\end{figure}

\autoref{fig:overview} presents an overview of \textsc{Flex}, an inference system that combines a request-adaptive hybrid model with an optimized serving runtime through algorithm-system co-design.

\noindent\textbf{Request-adaptive hybrid model.} The right part of \autoref{fig:overview} shows the proposed hybrid model, which consists of a stack of adaptive hybrid blocks. Each block contains a lightweight gate, paired Attention and SSM branches, and shared components such as FFNs. For each request, the gates select one branch per block, forming a request-specific hybrid layout that balances modeling capability and inference cost.

\noindent\textbf{Serving runtime.} The left part of \autoref{fig:overview} presents the runtime, which addresses two challenges of dynamic hybrid execution: requests within a batch may select different branches, and requests sharing a textual prefix may require incompatible caches. The runtime incorporates two modules. The selective batching engine executes shared operations on the original packed batch and splits requests into Attention and SSM sub-batches for branch-specific computation. Hybrid-aware context caching matches both textual prefixes and layer-selection histories to identify reusable Attention KV caches and SSM states.


\noindent\textbf{Workflow.} Given incoming requests, the runtime forms serving batches and identifies candidate cached prefixes. During prefill, the gate at each hybrid block selects one branch per request, and the runtime executes the selected branches while reusing caches compatible with the layer-selection history up to that block. The runtime records these choices and keeps them fixed throughout decoding, grouping requests accordingly and updating their Attention KV caches or SSM states until generation completes.

\section{Prefill-Oriented Gating}\label{sec:gating}

This section presents the prefill-oriented gating mechanism of \textsc{Flex}. The goal is to enable request-adaptive Attention and SSM layer selection while preserving efficient autoregressive decoding.

\subsection{Gating Design}
A natural way to enable dynamic hybrid execution is to make layer choices for each token. However, such token-level gating is ill-suited for autoregressive hybrid LLM serving. During decoding, each newly generated token relies on the Attention KV caches or SSM state caches produced by previous tokens to avoid redundant computation. If different tokens from the same request choose different layers within the same block, later tokens may no longer be able to directly reuse the caches produced by earlier tokens. As a result, the runtime must recompute previous tokens under the new layer choice, introducing substantial overhead and complicating decoding.


\begin{figure}[t]
    \centering
    \includegraphics[width=\linewidth]{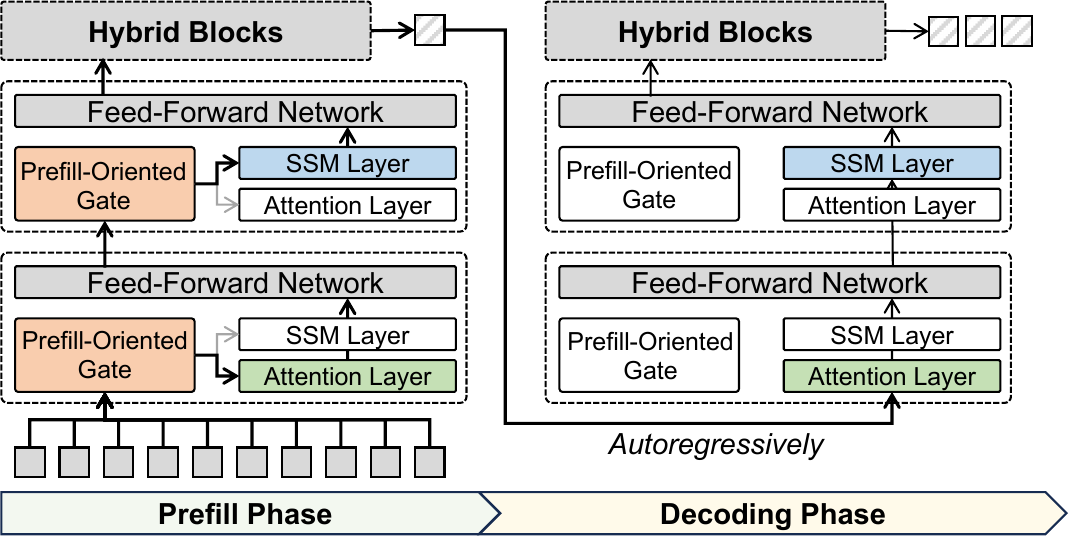}
    \caption{Illustration of the prefill-oriented gating mechanism. The activated layers are shown with filled colors, while inactive layers are left unfilled.}
    \label{fig:prefill_gate}
\end{figure}

To avoid these problems, \textsc{Flex} adopts a coarse-grained prefill-oriented gating mechanism in the request-adaptive hybrid model. As shown in \autoref{fig:prefill_gate}, this mechanism has two key properties. First, the gate is evaluated only during prefill, where each hybrid block makes one request-level choice between the Attention layer and the SSM layer based on all input tokens. Second, the selected layer choices are fixed after prefill and directly reused throughout decoding, so generated tokens do not trigger additional gating decisions. 

This design is reasonable for both system efficiency and hybrid adaptation. From the system perspective, prefill-oriented gating preserves a consistent computation throughout model inference and newly generated tokens can directly reuse the Attention KV caches or SSM state caches produced by previous tokens, avoiding the recomputation overhead. 
From the model perspective, the input tokens processed during prefill already encode the key characteristics of the request, such as task type, context length, and dependency requirements~\cite{zheng2023response, qiu2024efficient}. Our experimental results further show that such prefill information is sufficient to guide the Attention and SSM layer choices for the whole request, enabling the hybrid model to adapt its computation without repeatedly invoking the gate during decoding.

In each hybrid block, the gate first aggregates the prefill hidden states into a compact request representation, and then feeds it into a lightweight gate network to obtain the probabilities of selecting the Attention and SSM layers. We use top-1 gating and activate the layer with the larger probability. The formal definition of the gating process is provided in the supplementary manuscript.


\subsection{Hybrid Model Training}
The hybrid model in \textsc{Flex} is initialized from pretrained Attention and SSM components~\cite{dong2025hymba, li2025transmamba}. We use structured pruning to align their hidden dimensions~\cite{muralidharan2024compact}. When their depths differ, we select source layers using KL-divergence-based importance scores and pair the selected Attention and SSM layers to form hybrid blocks. We jointly fine-tune the gates and paired branches while freezing the shared embedding, FFN, and normalization modules initialized from the Attention backbone.

To ensure fine-tuning quality, we use a soft routing mode to keep optimization differentiable. For the $\ell$-th hybrid block, let $H^{\ell-1}$ denote the hidden states from the previous block, and let $\mathrm{Attn}^{\ell}(\cdot)$ and $\mathrm{SSM}^{\ell}(\cdot)$ denote the candidate Attention and SSM layers in this block, respectively. The gate outputs the probabilities of selecting the Attention and SSM layers. Instead of activating only one layer, the block computes a weighted combination of the two branch outputs:
\begin{align}
    \widetilde{H}^{\ell} = p_{\mathrm{attn}}^{\ell} \mathrm{Attn}^{\ell}(H^{\ell-1})
    +
    p_{\mathrm{ssm}}^{\ell} \mathrm{SSM}^{\ell}(H^{\ell-1}),
\end{align}
where $p_{\mathrm{attn}}^{\ell}$ and $p_{\mathrm{ssm}}^{\ell}$ are the gate probabilities for the Attention and SSM layers at the $\ell$-th hybrid block, and $\widetilde{H}^{\ell}$ is the output before the following feed-forward network. This soft routing formulation allows gradients to update the gate as well as both candidate layers. During inference, the model switches to hard routing, where each hybrid block activates only the layer with the larger gate probability.

The training objective balances generation quality and inference efficiency. For generation quality, we use the standard supervised fine-tuning loss. To encourage efficient layer selection, we add a lightweight cost-aware regularization term:
\begin{align}
    \mathcal{L}=
    \mathcal{L}_{\mathrm{sft}} +
    \lambda\frac{1}{N}\sum_{\ell=1}^{N}
    \left(p_{\mathrm{attn}}^{\ell} \bar{c}_{\mathrm{attn}}
    +
    p_{\mathrm{ssm}}^{\ell} \bar{c}_{\mathrm{ssm}}\right),
\end{align}
where $\mathcal{L}_{\mathrm{sft}}$ denotes the standard supervised fine-tuning loss for next-token prediction, $N$ is the number of hybrid blocks, and $\bar{c}_{\mathrm{attn}}$ and $\bar{c}_{\mathrm{ssm}}$ are normalized layer-level cost weights for selecting the Attention and SSM layers, respectively. The coefficient $\lambda$ controls the trade-off between generation quality and inference efficiency. This regularization is used only to bias the gate toward efficient layer choices during training.

\section{Selective Batching Engine}\label{sec:batching}
This section introduces the design of selective batching engine in \textsc{Flex}, which supports efficient batching processing of the proposed request-adaptive hybrid model.

\subsection{Split-and-Merge Execution}
\begin{figure}[t]
    \centering
    \includegraphics[width=\linewidth]{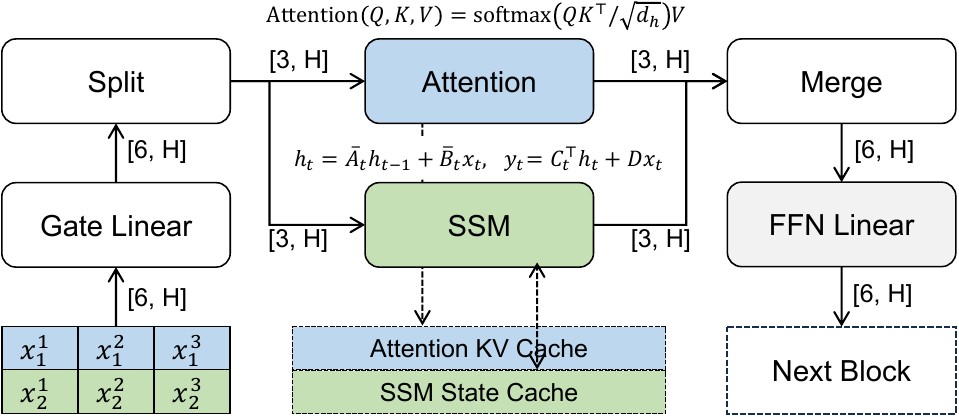}
    \caption{Illustration of the proposed split-and-merge mechanism in selective batching engine. We ignore \textit{Layer Norm}, \textit{Add}, and other operations for simplicity.}
    \label{fig:engine}
\end{figure}
Given the request-adaptive layer choices produced by dynamic hybrid execution, the key question is how to execute a mixed batch without sacrificing normal batching efficiency. \textsc{Flex} addresses this through a split-and-merge execution mechanism. As shown in \autoref{fig:engine}, the input tokens from different requests are first packed into a single batch tensor. For example, $x_i^j$ denotes the $j$-th token of request $i$, and the two requests in the figure form a packed tensor with shape $[6,H]$. \textsc{Flex} keeps this packed layout for shared operations, such as gate network and FFN computation, and splits the batch only before the layer-specific Attention or SSM computation. Requests selecting the Attention layer are compacted into an Attention sub-batch, while requests selecting the SSM layer are compacted into an SSM sub-batch. The two sub-batches are then executed by their corresponding kernels. Afterward, \textsc{Flex} merges their outputs back into the original packed batch layout, so subsequent shared operations can still be processed as a regular batch.

This design confines dynamic layer selection to the Attention and SSM computation inside each hybrid block. Other operations still use the standard packed tensor interface and therefore preserve normal batching efficiency. During prefill, the gate of each hybrid block is executed once to determine whether each request uses the Attention or SSM layer, and the selected layer processes all input tokens of that request. Meanwhile, the corresponding Attention KV or SSM state is cached. After prefill, each request records its selected layer choices across all hybrid blocks. During decoding, the gate is disabled, and \textsc{Flex} directly reuses the recorded layer choices to form Attention and SSM sub-batches at each block. The Attention layer reads and appends the Attention KV cache, while the SSM layer reads and updates the SSM cache.

\subsection{Load-Aware Branch Coordination}
Although selective batching enables the Attention and SSM layers to run independently, simply launching them on normal CUDA streams is insufficient. The two branches have different execution characteristics: the Attention branch is sensitive to context length and KV access, while the SSM branch is mainly dominated by recurrent state updates. Moreover, the number of tokens assigned to each branch changes across blocks and batches. As a result, one branch may finish earlier than the other, forcing the merge operation to wait for the slower branch.

To reduce this synchronization stall, \textsc{Flex} adopts load-aware branch coordination. The key idea is to estimate the relative workload of the Attention and SSM branches and assign more SM resources to the heavier branch. Guided by the computation complexity in \autoref{tab:complexity}, \textsc{Flex} uses lightweight profiled cost models to predict branch execution costs from simple runtime features, including the number of requests in each branch, input lengths during prefill, and current KV-cache lengths during decoding. Since branch coordination only requires relative costs, the predictor does not need exact latency estimation. In our implementation, its prediction error is within 6.32\% for the Attention branch and 5.18\% for the SSM branch. The profiling is a one-time effort for each model-GPU pair and can be completed efficiently. Detailed cost models are provided in the supplementary manuscript.

\begin{table}[t]
    \centering
    \label{tab:attn_ssm_complexity}
    \begin{tabular}{c|cc}
        \toprule
        & \textbf{Prefill} & \textbf{Decoding} \\
        \midrule
        \textbf{Attention Layer} 
        & $\mathcal{O}(Ld^2 + L^2d)$ 
        & $\mathcal{O}(d^2 + rd)$ \\
        \textbf{SSM Layer} 
        & $\mathcal{O}(Ld^2 + Ld_s)$ 
        & $\mathcal{O}(d^2 + d_s)$ \\
        \bottomrule
    \end{tabular}
    \caption{Computation complexities of Attention and SSM layers. Here, $L$ denotes the input length in prefill, $r$ denotes the current cached context length during decoding, $d$ denotes hidden dimension, and $d_s$ denotes SSM state dimension.}
    \label{tab:complexity}
\end{table}


Based on the predicted costs, \textsc{Flex} partitions the schedulable GPU resources between the Attention and SSM sub-batches at the SM granularity. In our implementation, we use NVIDIA H100 SXM GPUs and empirically set the allocation granularity to $G=16$ SMs, following recent intra-GPU multiplexing practice~\cite{gao2025duetserve, zhu2025nanoflow, chen2026towards}. Of the 132 SMs available on each GPU, 128 are included in the partitioned resource pool, yielding a compact set of feasible allocations in multiples of 16 SMs. \textsc{Flex} rounds the target allocation to the nearest feasible partition and assigns the remaining schedulable SMs to the other sub-batch. If only one sub-batch is nonempty, it skips partitioning and assigns the entire resource pool to that sub-batch. This compact allocation space enables lightweight online selection while providing sufficient flexibility to balance the two computations. The detailed SM allocation rule is provided in the supplementary manuscript.

The selected SM partition is implemented using NVIDIA GreenContext~\cite{greencontext}. \textsc{Flex} creates separate Green Contexts and CUDA streams for the Attention and SSM sub-batches, binding each stream to its assigned SM partition. Since GreenContext only constrains the SM scheduling pool, existing FlashAttention and \code{mamba\_ssm} kernels can run without modification. The two sub-batches are launched concurrently and synchronized with CUDA events before the merge operation and the next hybrid block. This implementation assigns more compute resources to the heavier sub-batch and reduces the completion-time imbalance between them.

\section{Hybrid-Aware Context Caching}\label{sec:context}
This section presents the hybrid-aware context caching mechanism in \textsc{Flex}, which enables correct prefix reuse when the same textual context may produce different layer-wise Attention KV and SSM state caches.
\begin{figure}[t]
    \centering
    \includegraphics[width=\linewidth]{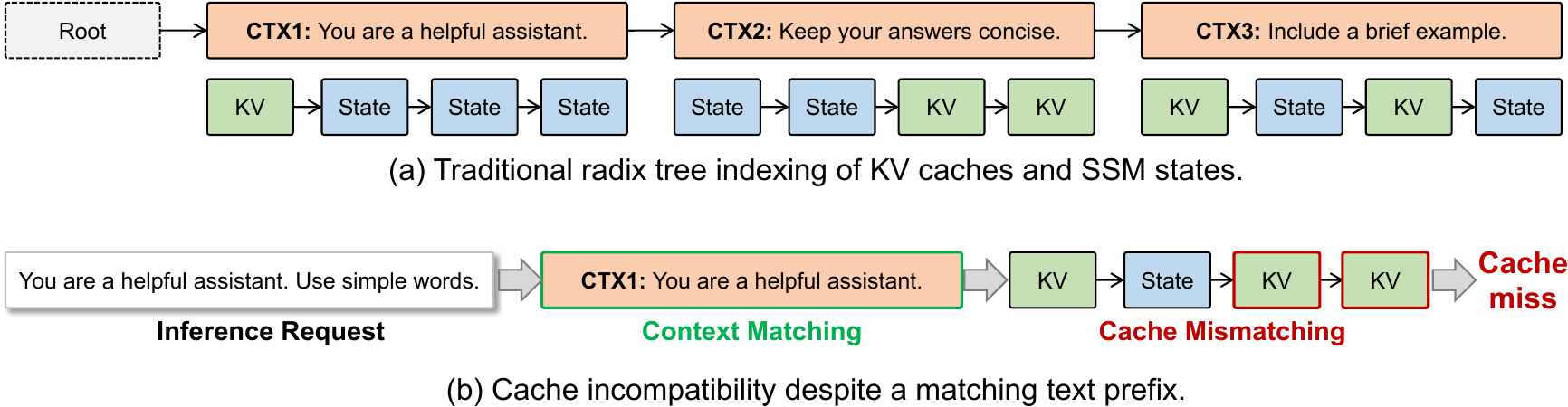}
    \caption{Motivation for hybrid-aware context caching. (a) A conventional radix tree indexes Attention KV caches and SSM states by textual prefixes. (b) The incoming request matches context 1 (CTX1) but selects different layer types, resulting in a cache miss.}
    \label{fig:radix_example}
\end{figure}

\subsection{Context Caching Mechanism}
Context caching is widely used to improve LLM serving efficiency by reusing previously computed caches (e.g., Attention KVs and SSM states) for repeated prefixes~\cite{jin2025ragcache, yao2025cacheblend, yu2025ic}. For example, RadixAttention~\cite{zheng2024sglang} organizes historical contexts with a radix tree, where each node represents a textual prefix. When a new request arrives, the runtime searches the radix tree to find the longest matched prefix, reuses the caches associated with the matched prefix, and computes only the unmatched suffix. However, conventional RadixAttention cannot be directly applied to \textsc{Flex}. Its key assumption is that a matched textual prefix uniquely determines all layer-wise caches that can be reused. This assumption holds for static models because the layer type at each position is fixed by the model architecture. For example, in an Attention-only model, each layer always produces a KV cache, while in an SSM-only model, each layer always produces an SSM cache. 

In \textsc{Flex}, requests sharing a textual prefix may select different layer layouts depending on their full contexts. As shown in \autoref{fig:radix_example}, the incoming request matches the cached prefix ``You are a helpful assistant.'', but requires Attention KV caches at the third and fourth blocks, where the cached entries are SSM states. Conventional methods treat the matched entry as a whole: it is either reused or rejected. Reusing the cached states despite the layout mismatch can introduce inconsistent intermediate representations and degrade inference quality. Alternatively, rejecting the incompatible entry results in a cache miss and requires recomputing the shared prefix, losing the opportunity to exploit the textual overlap.

To support correct and efficient context reuse under dynamic hybrid execution, \textsc{Flex} implements hybrid-aware context caching using a two-level radix-tree structure. The outer-context radix tree indexes textual prefixes, while an inner-cache radix tree attached to each context node organizes layer-selection paths and their associated Attention KV caches and SSM states. Cache lookup proceeds in two steps. First, \textsc{Flex} locates the longest matching textual prefix in the outer-context radix tree. It then examines the inner-cache radix tree of the matched node and its ancestors to identify compatible cache paths. Compatibility requires matching the sequence of layer choices from the first block through the block being reused. Among these candidates, \textsc{Flex} selects the prefix and cache path with the lowest estimated recomputation cost, considering both prefix length and the number of reusable leading blocks.

\begin{figure}[t]
    \centering
    \includegraphics[width=\linewidth]{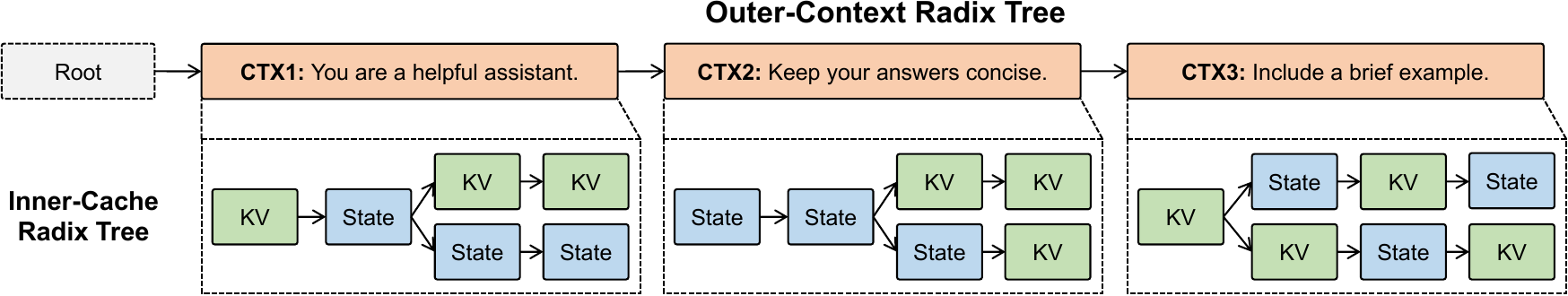}
    \caption{Illustration of the proposed two-level radix-tree structure. The outer-context radix tree indexes contextual prefixes, while each inner-cache radix tree organizes alternative layer-selection paths and their associated Attention KV and SSM state caches, enabling cache reuse along compatible execution paths for a matched prefix.}
    \label{fig:2dredixtree}
\end{figure}

As illustrated in \autoref{fig:2dredixtree}, the inner-cache radix tree associated with CTX1 contains multiple execution paths. These paths share caches for the first two blocks, then branch according to the branch choices at the third and fourth blocks. For cache, \textsc{Flex} retains input hidden states and terminal SSM states, reconstructing Attention KV caches on demand. For an incoming request matching CTX1, \textsc{Flex} follows the inner-tree path consistent with its branch-selection history and reuses the corresponding caches. At the first route mismatch, the cached hidden states remain valid because all preceding branch choices match. \textsc{Flex} uses these states to recompute the prefix at that block and all subsequent blocks, then extends the inner-cache radix tree with the newly computed caches while retaining existing branches. This organization supports partial cache reuse and preserves alternative execution paths for future requests sharing same textual prefix.


\subsection{Priority-based Eviction Policy}
Hybrid-aware context caching enables correct cache reuse under dynamic hybrid execution, but also increases the number of cache entries. With a fixed layer layout, conventional RadixAttention associates each textual prefix with one set of layer-wise caches. In \textsc{Flex}, the same context node can maintain multiple cache paths corresponding to different Attention and SSM choices. Evicting an entire context node therefore discards all these paths, potentially removing frequently reused caches together with rarely accessed ones.

\textsc{Flex} instead performs eviction at the granularity of individual layer-wise cache entries. Under GPU memory pressure, it evicts entries from the inner-cache radix trees. An outer-context node is removed only when all its associated caches have been evicted and the node is no longer needed to represent shared textual prefixes. To determine which entries to retain, \textsc{Flex} assigns each cache $c$ a retention priority:
\begin{align}
    P(c) =\frac{H(c) \cdot R(c)}{M(c)} \cdot D(c),
\end{align}
where $H(c)$ is its recent access frequency, $R(c)$ is the recomputation cost saved by reusing it, $M(c)$ is its memory footprint, and $D(c)$ is a depth factor that favors caches at earlier blocks. \textsc{Flex} evicts entries with the lowest priority first. This policy favors caches that provide greater expected computation savings per unit of memory. The depth factor further prioritizes early-layer caches, which can form shared ancestors of multiple execution paths in an inner-cache radix tree.

\section{Implementation}
We implement a prototype of \textsc{Flex} with approximately 6.5K lines of code using Python, PyTorch, HuggingFace Transformers, and CUDA/C++. The construction and fine-tuning of the request-adaptive hybrid model are implemented in PyTorch, while the serving runtime extends a vLLM-style execution loop with selective batching and hybrid-aware context caching. It caches layer hidden states and terminal SSM states, supporting Attention KV reconstruction and recomputation from the first mismatched layer. For concurrent Attention and SSM execution, \textsc{Flex} uses NVIDIA GreenContext to partition SM resources and launches the two sub-batches on separate CUDA streams. It also adopts a unified paging-based memory manager for heterogeneous Attention KV caches and SSM caches, following Jenga~\cite{zhang2025jenga}. The runtime currently uses a conventional continuous batching scheduler~\cite{yu2022orca}, and is compatible with other scheduling policies such as chunked prefill, prefill-decode disaggregation, and latency-aware scheduling~\cite{agrawal2024taming,du2025prefillonly,patel2024splitwise,zhong2024distserve,qiu2024efficient}. Detailed implementation is provided in the supplementary manuscript.

\section{Evaluation}
\subsection{Evaluation Setup}
\noindent\textbf{Platform.} We evaluate \textsc{Flex} on a server equipped with one NVIDIA H100 SXM GPU (80GB memory) and Intel Xeon Platinum 8458P CPUs. The server runs Ubuntu 22.04.5 LTS with NVIDIA driver 580.82.07 and CUDA runtime 12.4. We use Python 3.12.3, PyTorch 2.4.1, Transformers 4.46.0 and mamba\_ssm 2.2.4.

\noindent\textbf{Models and baselines.} We evaluate the proposed hybrid model against both homogeneous and static hybrid LLMs. The homogeneous baselines include Llama-3~\cite{grattafiori2024llama} as an Attention-only model and Mamba-2~\cite{dao2024transformers} as an SSM-only model. The static hybrid baselines include Zamba~\cite{glorioso2024zamba2}, Jamba~\cite{lenz2025jamba}, and Nemotron-H~\cite{blakeman2025nemotron}, whose Attention-to-SSM layouts are fixed after model construction. For serving-system evaluation, we compare \textsc{Flex} with vLLM~\cite{kwon2023efficient} and SGLang~\cite{zheng2024sglang}, two widely used LLM serving systems, as well as Jenga~\cite{zhang2025jenga}, which further provides memory management support for Attention KV and SSM states. All systems use the same model, workload, and generation settings, with unrelated optimizations such as speculative decoding disabled.



\noindent\textbf{Datasets and workloads.} We fine-tune each model on a 10K-example subset of ShareGPT-90K and reuse the resulting checkpoint across all evaluation tasks without task-specific fine-tuning. We evaluate model quality on six datasets covering reasoning, question answering (QA), and summarization. For reasoning, we use ARC-Easy and ARC-Challenge~\cite{clark2018think}. For QA, we use PIQA~\cite{bisk2020piqa} and TruthfulQA~\cite{lin2022truthfulqa}. For summarization, we use CNN/DailyMail~\cite{cnndm} and PubMed~\cite{cohan2018discourse}.

\noindent\textbf{Metrics.}
For model quality, we report accuracy for ARC-Easy, ARC-Challenge, and PIQA, MC1 accuracy for TruthfulQA, and ROUGE-L for CNN/DM and PubMed. For model efficiency, we estimate decoding TFLOPs per token:
\begin{align}
\mathrm{TFLOPs/token}\!=\!\frac{F^{\mathrm{lin}}_{r,t}\!+\!4L_{r,t}\sum_{\ell\in\mathcal{A}_r}n_{q,\ell}d_{h,\ell}\!+\!\sum_{\ell\in\mathcal{S}_r}C^{\mathrm{SSM}}_\ell}{10^{12}},
\end{align}
where $F^{\mathrm{lin}}_{r,t}$ counts executed linear operations, including projections, FFNs, and the output head. $\mathcal{A}_r$ and $\mathcal{S}_r$ are the Attention and SSM layers activated by request $r$. $L_{r,t}$ is the attended KV length at decoding step $t$,  and $n_{q,\ell}$ and $d_{h,\ell}$ are the query-head count and head dimension. For \textsc{Flex}, only the selected branch at each block is counted, together with the shared operations. $C^{\mathrm{SSM}}_\ell$ counts convolution, state update, and readout operations beyond the linear projections~\cite{gu2023mamba, dao2024transformers}. We count each multiply-add as two FLOPs~\cite{hoffmann2022training, chowdhery2023palm}, omit normalization, activations, and softmax, and average over evaluated decoding steps. For serving performance, we report request throughput, time-to-first-token (TTFT), time-per-output-token (TPOT), and end-to-end latency.

\noindent\textbf{Other configurations.} Fine-tuning uses Adam with a learning rate of $2\times10^{-4}$, an effective batch size of 64, and a maximum sequence length of 4,096 tokens. We train for 1,000 optimization steps using BF16 precision and fully sharded data parallelism (FSDP), with $\lambda=0.2$ across all model scales and benchmarks.

\begin{table*}[t]
\centering
\small
\setlength{\tabcolsep}{5pt}
\renewcommand{\arraystretch}{1.08}
\resizebox{\textwidth}{!}{
\begin{tabular}{l|ccccc|cc|cc|cc|c}
\toprule
\multirow{2}{*}{\textbf{Models}}
& \multicolumn{5}{c|}{\textbf{Base Config}}
& \textbf{ARC-E}
& \textbf{ARC-C}
& \textbf{PIQA}
& \textbf{TruthfulQA}
& \textbf{CNN/DM}
& \textbf{PubMed}
& \textbf{Model Efficiency} \\
\cmidrule(lr){2-6}
\cmidrule(lr){7-8}
\cmidrule(lr){9-10}
\cmidrule(lr){11-12}
\cmidrule(lr){13-13}
& $P$
& $P_{\mathrm{act}}$
& $N_{\mathrm{attn}}$
& $N_{\mathrm{ssm}}$
& $H_{\mathrm{hidden}}$
& Accuracy $\uparrow$
& Accuracy $\uparrow$
& Accuracy $\uparrow$
& Accuracy (MC1) $\uparrow$
& ROUGE-L $\uparrow$
& ROUGE-L $\uparrow$
& TFLOPs/Token \\
\midrule

Mamba
& 5.3B & 5.3B & -- & 64 & 2560
& 67.4 $\pm$ 0.11
& 35.8 $\pm$ 0.13
& 74.9 $\pm$ 0.14
& 33.3 $\pm$ 0.07
& 10.9 $\pm$ 0.14
& 14.3 $\pm$ 0.15
& \num{1.01e-02} \\

Llama
& 3.2B & 3.2B & 28 & -- & 3072
& 73.2 $\pm$ 0.08
& 42.4 $\pm$ 0.12
& 75.5 $\pm$ 0.15
& 42.1 $\pm$ 0.33
& 17.1 $\pm$ 0.27
& 15.9 $\pm$ 0.18
& \num{1.08e-02} \\

Zamba
& 2.7B & 2.7B & 9 & 45 & 2560
& 69.4 $\pm$ 0.24
& 41.9 $\pm$ 0.25
& \textbf{\underline{78.7 $\pm$ 0.31}}
& 41.8 $\pm$ 0.17
& \textbf{\underline{20.5 $\pm$ 0.22}}
& 16.0 $\pm$ 0.19
& \num{7.59e-03} \\

Jamba
& 3.0B & 3.0B & 2 & 26 & 2560
& 69.7 $\pm$ 0.15
& \textbf{\underline{47.3 $\pm$ 0.26}}
& 70.5 $\pm$ 0.32
& 48.8 $\pm$ 0.16
& 18.9 $\pm$ 0.33
& 18.1 $\pm$ 0.19 & \num{6.02e-03} \\

Nemotron-H
& 4.5B & 4.5B & 4 & 24 & 3072
& 74.0 $\pm$ 0.24
& 44.6 $\pm$ 0.17
& 71.1 $\pm$ 0.41
& 49.9 $\pm$ 0.33
& 16.8 $\pm$ 0.19
& 16.5 $\pm$ 0.20 & \num{8.98e-03} \\

\rowcolor{gray!20}
\textsc{Ours}
& 4.4B & 3.2B & 28 & 28 & 2560
& \textbf{\underline{74.7 $\pm$ 0.17}}
& 45.5 $\pm$ 0.26
& 76.6 $\pm$ 0.31
& \textbf{\underline{51.4 $\pm$ 0.19}}
& 20.1 $\pm$ 0.25
& \textbf{\underline{19.4 $\pm$ 0.24}} & \num{6.43e-03} \\
\bottomrule

Mamba
& 8.2B & 8.2B & -- & 56 & 4096
& 72.5 $\pm$ 0.23
& 50.9 $\pm$ 0.19
& 76.6 $\pm$ 0.33
& 38.8 $\pm$ 0.32
& 16.7 $\pm$ 0.24
& 15.3 $\pm$ 0.25 & \num{1.71e-02} \\

Llama
& 13.1B & 13.1B & 55 & -- & 4096
& 76.7 $\pm$ 0.17
& 54.3 $\pm$ 0.17
& 79.5 $\pm$ 0.19
& 59.8 $\pm$ 0.24
& 19.5 $\pm$ 0.10
& 16.4 $\pm$ 0.22 & \num{2.62e-02} \\

Zamba
& 7.4B & 7.4B & 13 & 68 & 3584
& 70.3 $\pm$ 0.30
& 45.4 $\pm$ 0.17
& 81.1 $\pm$ 0.22
& 65.6 $\pm$ 0.24
& 20.1 $\pm$ 0.32
& 16.2 $\pm$ 0.15 & \num{1.98e-02} \\

Jamba
& 13.0B & 13.0B & 1 & 7 & 4096
& 72.8 $\pm$ 0.24
& 52.5 $\pm$ 0.22
& 80.3 $\pm$ 0.38
& 66.2 $\pm$ 0.41
& 19.4 $\pm$ 0.15
& 18.3 $\pm$ 0.18 & \num{2.68e-02} \\

Nemotron-H
& 12.3B & 12.3B & 6 & 28 & 5120
& 77.3 $\pm$ 0.18
& 56.1 $\pm$ 0.17
& 82.2 $\pm$ 0.26
& 68.7 $\pm$ 0.41
& 18.3 $\pm$ 0.39
& 16.8 $\pm$ 0.25 & \num{2.46e-02} \\

\rowcolor{gray!20}
\textsc{Ours}
& 14.5B & 6.5B & 32 & 32 & 4096
& \textbf{\underline{77.8 $\pm$ 0.17}}
& \textbf{\underline{57.7 $\pm$ 0.36}}
& \textbf{\underline{83.1 $\pm$ 0.28}}
& \textbf{\underline{68.9 $\pm$ 0.39}}
& \textbf{\underline{23.4 $\pm$ 0.14}}
& \textbf{\underline{21.5 $\pm$ 0.13}} & \num{1.75e-02} \\
\bottomrule

Mamba
& -- & -- & -- & -- & --
& --
& --
& --
& --
& --
& -- & --\\

Llama
& 70B & 70B & 80 & -- & 8192
& 84.3 $\pm$ 0.16
& 82.7 $\pm$ 0.24
& 84.3 $\pm$ 0.33
& 63.7 $\pm$ 0.18
& \textbf{\underline{25.7 $\pm$ 0.09}}
& \textbf{\underline{23.6 $\pm$ 0.11}} & \num{1.41e-01} \\

Zamba
& -- & -- & -- & -- & --
& --
& --
& --
& --
& --
& -- & -- \\

Jamba
& 52B & 12B & 4 & 28 & 4096
& 73.7 $\pm$ 0.22
& 63.6 $\pm$ 0.19
& 83.1 $\pm$ 0.31
& 72.1 $\pm$ 0.17
& 20.5 $\pm$ 0.16
& 19.8 $\pm$ 0.11 & \num{2.40e-02} \\

Nemotron-H
& 56B & 56B & 10 & 54 & 8192
& 85.2 $\pm$ 0.23
& 84.8 $\pm$ 0.17
& 84.4 $\pm$ 0.12
& 73.4 $\pm$ 0.26
& 20.1 $\pm$ 0.19
& 17.9 $\pm$ 0.28 & \num{1.12e-01} \\

\rowcolor{gray!20}
\textsc{Ours}
& 61B & 23B & 32 & 32 & 8192
& \textbf{\underline{85.6 $\pm$ 0.21}}
& \textbf{\underline{85.1 $\pm$ 0.17}}
& \textbf{\underline{85.3 $\pm$ 0.25}}
& \textbf{\underline{75.2 $\pm$ 0.31}}
& 24.5 $\pm$ 0.27
& 22.3 $\pm$ 0.21 & \num{4.63e-02} \\
\bottomrule

\end{tabular}
}
\caption{Comparison of model quality and efficiency across homogeneous, static hybrid, and our proposed request-adaptive hybrid model. $P$ and $P_{act}$ denote the total model parameters and the activated parameters for one request, respectively. $N_{\mathrm{attn}}$ and $N_{\mathrm{ssm}}$ denote the numbers of Attention and SSM layers. For \textsc{Flex}, $P$ counts all paired candidate branches, while $P_{\mathrm{act}}$ reports the average activated parameters under request-adaptive layer selection. All multiple-choice tasks use free-form generation with regex-based letter extraction. Chat-template-aware models receive a ``letter only'' system prompt and reasoning models use \texttt{no\_think} mode.}
\label{tab:model_performance}
\end{table*}
\subsection{Overall Model Performance}
\noindent\textbf{Quality-efficiency trade-off.} \autoref{tab:model_performance} compares \textsc{Flex} with homogeneous and static hybrid LLMs across three model scales. Overall, \textsc{Flex} achieves the best results on 13 out of 18 quality metrics while maintaining competitive computation cost. At the small scale, \textsc{Flex} obtains the best results on ARC-Easy, TruthfulQA, and PubMed, and remains close to the best baseline on ARC-Challenge and CNN/DailyMail. At the medium scale, it consistently outperforms all baselines on all six datasets. At the large scale, it achieves the best results on reasoning and QA tasks, while remaining competitive on summarization tasks. More importantly, \textsc{Flex} achieves these quality gains with substantially lower estimated TFLOPs/token: compared with Attention-only Llama models, it reduces computation cost by 40.5\%, 33.2\%, and 67.2\% at the three evaluated scales, respectively. Compared with Nemotron-H, it reduces TFLOPs/token by 28.4\%, 28.9\%, and 58.7\%. Although Mamba and some static hybrid models can be cheaper in certain settings, their quality is often less stable across tasks. These results show that request-adaptive Attention and SSM selection provides a better quality-efficiency trade-off than static hybrid layouts.
We further analyze evaluation results across different tasks.

\noindent\textbf{Reasoning tasks.} On ARC-Easy and ARC-Challenge, \textsc{Flex} performs particularly well at medium and large scales. For example, at the medium scale, \textsc{Flex} achieves 77.8\% accuracy on ARC-Easy and 57.7\% on ARC-Challenge, outperforming all static hybrid baselines. At the large scale, it further improves the results to 85.6\% and 85.1\%, respectively. These results suggest that reasoning tasks benefit from selectively activating Attention in important layers, while still using efficient SSM computation elsewhere. Static hybrid models improve over pure SSM models, but their static attention placement cannot adapt to the varying dependency patterns of different requests.

\noindent\textbf{QA tasks.} On PIQA and TruthfulQA, \textsc{Flex} also shows strong performance. At the medium scale, it achieves 83.1\% on PIQA and 68.9\% on TruthfulQA, outperforming all baselines. At the large scale, it reaches 85.3\% on PIQA and 75.2\% on TruthfulQA. These results indicate that QA workloads do not always require Attention in every layer, but still need Attention in selected layers to preserve factual and contextual signals. \textsc{Flex} captures this behavior by adapting the Attention and SSM choices to each request instead of relying on a static hybrid layout.

\noindent\textbf{Summarization tasks.} Summarization workloads have a stronger dependence on global context modeling. Attention-heavy models such as Llama remain competitive, especially at the large scale. Nevertheless, \textsc{Flex} achieves the best summarization results at the medium scale, reaching 23.4 ROUGE-L on CNN/DailyMail and 21.5 ROUGE-L on PubMed. At the small scale, it also achieves the best PubMed score and remains close to the best CNN/DailyMail result. These results show that summarization benefits from more frequent Attention activation, but does not require all layers to be Attention layers. \textsc{Flex} can allocate Attention to the layers and requests that need global context, while using SSM layers elsewhere to reduce computation.

\begin{table}[t]
\centering
\small
\resizebox{.475\textwidth}{!}{
\begin{tabular}{c|l|cc|cc|cc}
\toprule
\textbf{Scale}
& \textbf{Method}
& \textbf{ARC-E}
& \textbf{ARC-C}
& \textbf{PIQA}
& \textbf{TruthfulQA}
& \textbf{CNN/DM}
& \textbf{PubMed} \\
\midrule

\multirow{3}{*}{4.4B}
& G-Optimal
& 68.2 & 37.2 & 70.5 & 45.4 & 16.6 & 15.2 \\

& D-Optimal
& 71.3 & 40.4 & 73.6 & 47.3 & 18.3 & 16.5 \\

& Adaptive
& 74.5
& 45.6
& 76.4
& 51.3
& 20.2
& 19.6 \\
\midrule

\multirow{3}{*}{14.5B}
& G-Optimal
& 71.8 & 49.5 & 73.7 & 60.3 & 17.6 & 18.8 \\

& D-Optimal
& 73.4 & 52.6 & 77.2 & 63.5 & 19.2 & 20.6 \\

& Adaptive
& 77.9
& 57.7
& 83.2
& 68.8
& 23.4
& 21.6 \\
\midrule

\multirow{3}{*}{61B}
& G-Optimal
& 74.3 & 72.1 & 75.4 & 69.7 & 20.3 & 19.6 \\

& D-Optimal
& 79.6 & 76.4 & 80.3 & 72.6 & 21.5 & 20.1 \\

& Adaptive
& 85.4
& 85.0
& 85.3
& 75.4
& 24.6
& 22.1 \\
\bottomrule
\end{tabular}
}
\caption{Comparison of routing methods across model scales.}
\label{tab:routing_benefits}
\end{table}
\noindent\textbf{Routing Effectiveness.} We compare three routing strategies using the same fine-tuned checkpoint at each model scale. We collect candidate layouts from adaptive routing on held-out validation data and evaluate each as a fixed layout under matched inference-cost budgets. Within this candidate set, \textit{Global Optimal} (G-Optimal) selects one layout that maximizes the macro-average of normalized validation scores across datasets, while \textit{Dataset Optimal} (D-Optimal) selects the best layout separately for each dataset. Both strategies keep their selected layouts fixed during evaluation, whereas Adaptive selects a layout for each request. As shown in \autoref{tab:routing_benefits}, our adaptive method outperforms both static strategies across all 18 model-dataset combinations. Compared with D-Optimal, it achieves average gains of 4.86 percentage points in accuracy and 2.55 points in ROUGE-L. Although D-Optimal consistently improves over G-Optimal, its remaining gap to Adaptive demonstrates the benefit of adapting layer choices to individual requests within the same dataset.

\subsection{Overall System Performance}
\begin{figure*}[t]
    \centering
    \includegraphics[width=\linewidth]{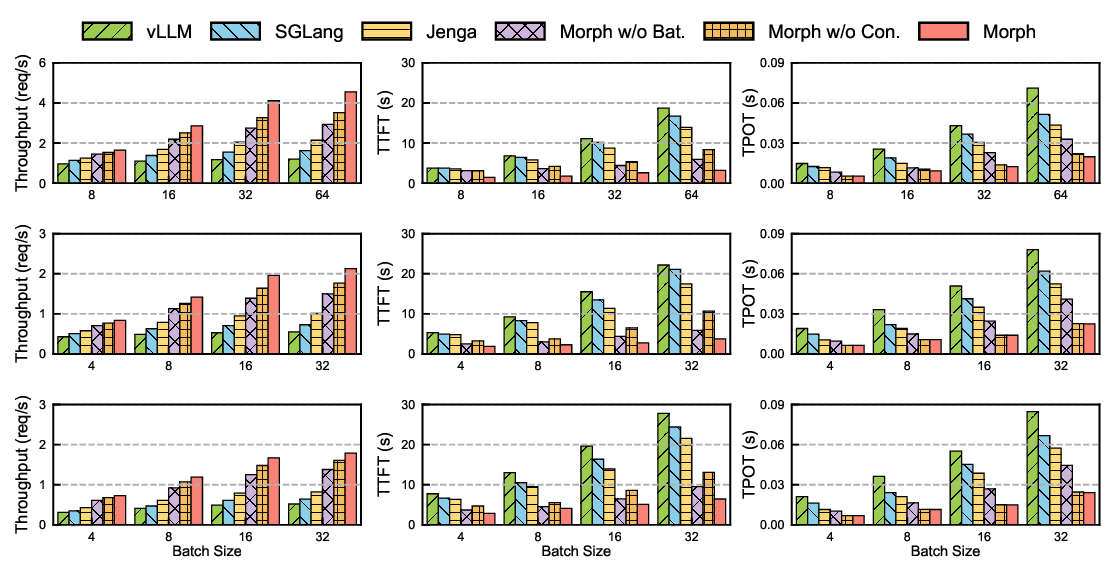}
    \caption{End-to-end serving performance of \textsc{Flex} under varying batch sizes on a single H100 GPU. Rows correspond to the 4.4B, 14.5B, and 61B models, respectively. Columns report request throughput, TTFT, and TPOT. The 61B model uses FP8 quantization to avoid out-of-memory.}
    \label{fig:overall_performance}
\end{figure*}
We evaluate the system-level performance of \textsc{Flex} at three model scales on a single H100 GPU, using FP8 quantization for the 61B model. We compare its serving runtime with vLLM, SGLang, and Jenga, with all methods serving the same model checkpoint at each scale. We also include two ablated variants: \textsc{Flex} w/o Bat. disables selective batching (\S\ref{sec:batching}), while \textsc{Flex} w/o Con. disables hybrid-aware context reuse (\S\ref{sec:context}). We construct mixed workloads from all datasets in which requests share article prefixes and append different questions. We vary the number of requests processed per scheduling iteration, using batch sizes of 8, 16, 32, and 64 for the 4.4B model and 4, 8, 16, and 32 for the other two scales. We report request throughput, TTFT, and TPOT.

\autoref{fig:overall_performance} shows that \textsc{Flex} consistently outperforms all three baselines in throughput, TTFT, and TPOT across the evaluated batch sizes and model scales. It achieves up to $3.79\times$, $3.87\times$, and $3.44\times$ the throughput of vLLM for the 4.4B, 14.5B, and 61B models, respectively. Its throughput advantage over vLLM grows with batch size. Larger batches offer more opportunities to group requests taking the same branch, which \textsc{Flex} exploits through selective batching. Both ablated variants also achieve lower throughput than the full system, showing that selective batching and context reuse contribute to serving efficiency.

Hybrid-aware context reuse substantially reduces TTFT on the mixed workload by avoiding repeated computation over shared article prefixes. Under dynamic layer choices, matching prefix tokens alone does not establish cache compatibility: the cached states also depend on the execution paths that produced them. \textsc{Flex} uses the two-level radix tree to check both textual prefixes and execution-path compatibility before reusing layer-wise Attention KV caches and SSM states. At the largest tested batch size for each scale, \textsc{Flex} reduces TTFT by an average of 80.9\%, 78.9\%, and 75.2\% compared with vLLM, SGLang, and Jenga, respectively. It also reduces TTFT by 59.1\% on average compared with \textsc{Flex} w/o Con., demonstrating the benefit of context reuse.

The TPOT results show the benefit of selective batching during decoding. Each request retains its layer choices after prefill, allowing \textsc{Flex} to group requests taking the same branch into compact Attention and SSM sub-batches. At the same configurations, \textsc{Flex} reduces TPOT by an average of 71.6\%, 63.1\%, and 56.5\% compared with vLLM, SGLang, and Jenga, respectively. Compared with \textsc{Flex} w/o Bat., it reduces TPOT by 43.6\% on average, demonstrating that selective batching improves decoding efficiency under request-specific layer choices.

\subsection{Ablation Study}
\begin{figure*}[t]
    \centering
    \includegraphics[width=\linewidth]{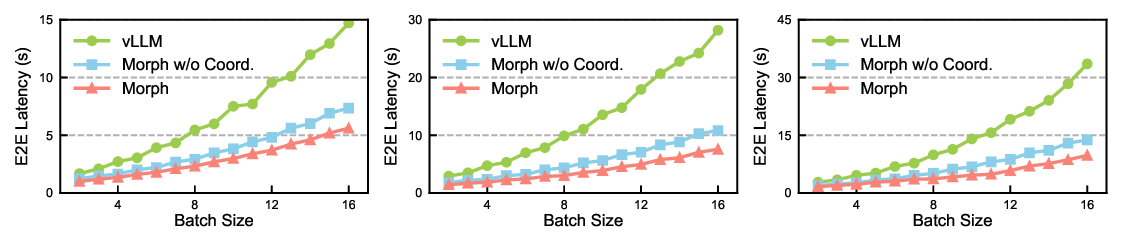}
    \caption{Ablation study of the selective batching engine and branch coordination under varying batch sizes. From left to right, the three subfigures report end-to-end latency for the 4.4B, 14.5B, and 61B hybrid models in \textsc{Flex}, respectively.}
    \label{fig:abl_batching}
\end{figure*}
\noindent\textbf{Effectiveness of selective batching.} We further evaluate the selective batching engine by varying the batch size from 2 to 16 on the PubMed dataset. We compare \textsc{Flex} with vLLM and an ablated variant, denoted as \textsc{Flex} w/o Coord., which retains split-and-merge batching but disables load-aware branch coordination for SM allocation between the Attention and SSM sub-batches. As shown in \autoref{fig:abl_batching}, \textsc{Flex} consistently achieves lower end-to-end (E2E) latency than vLLM across all tested batch sizes and model scales. Its selective batching engine groups requests according to their layer choices at each block, forming compact Attention and SSM sub-batches for execution with the corresponding kernels. Averaged over all batch sizes, \textsc{Flex} reduces E2E latency compared with vLLM by 54.6\%, 66.0\%, and 59.9\% for the 4.4B, 14.5B, and 61B models, respectively. At batch size 16, the corresponding reductions reach 61.9\%, 73.2\%, and 70.9\%, showing a larger latency advantage at the upper end of the tested batch sizes.

Comparing \textsc{Flex} with \textsc{Flex} w/o Coord. isolates the additional benefit of load-aware branch coordination. Although split-and-merge batching groups requests by their selected layers, the Attention and SSM sub-batches can still have imbalanced workloads, and the merge operation must wait for both branches to complete. \textsc{Flex} coordinates SM allocation according to branch workloads to reduce this imbalance. Compared with \textsc{Flex} w/o Coord., the full system reduces E2E latency by 21.7\%, 28.2\%, and 26.2\% on average for the three model scales, respectively. For the 61B model at batch size 16, vLLM requires 33.53s, while \textsc{Flex} w/o Coord. and the full \textsc{Flex} require 13.77s and 9.75s, respectively. Branch coordination thus provides a further 29.2\% latency reduction beyond split-and-merge batching alone in this configuration.

\begin{figure}[t]
    \centering
    \includegraphics[width=\linewidth]{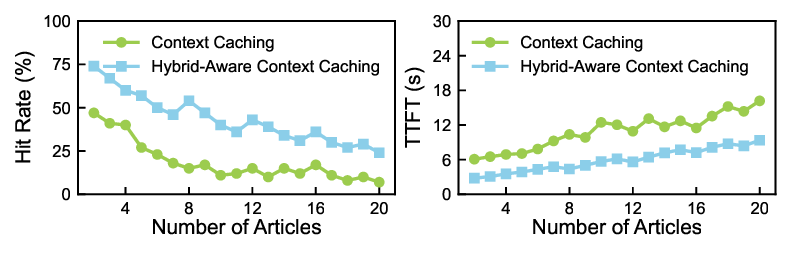}
    \caption{Effectiveness of hybrid-aware context caching under varying numbers of shared PubMed contexts.}
    \label{fig:abl_context}
\end{figure}
\noindent\textbf{Effectiveness of hybrid-aware context caching.} We evaluate the proposed hybrid-aware context caching mechanism using a PubMed-based shared-context workload. Each article serves as a shared textual prefix, and multiple requests append different questions after the same article~\cite{zhang2025jenga}. We vary the number of articles from 2 to 20 under a fixed cache capacity, increasing cache competition while preserving prefix reuse opportunities within each article. For conventional context caching~\cite{zheng2024sglang}, a context is counted as reusable only when all layer-wise caches associated with the matched prefix can be reused. If any Attention KV cache or SSM cache is incompatible with the current request's layer choices, the whole context is treated as a miss. In contrast, the proposed hybrid-aware context caching indexes layer-wise caches under each matched context, so it can reuse compatible caches even when later layers diverge. We compare the two mechanisms using effective cache hit rate and TTFT.

As shown in \autoref{fig:abl_context}, the hit rates of both methods decrease as the number of articles increases, because more shared contexts compete for the fixed memory capacity. Nevertheless, hybrid-aware context caching consistently maintains a higher effective hit rate. Across all settings, it achieves an average hit rate of 43.4\%, compared with 18.7\% for the baseline, improving the hit rate by 24.7 percentage points. When the number of articles reaches 20, conventional caching method retains only a 7\% hit rate, whereas hybrid-aware context caching still achieves 24\%. This confirms that layer-wise cache indexing captures valid reuse opportunities that conventional prefix-level caching would discard once any layer becomes incompatible.

The higher hit rate directly reduces prefill latency. As the number of articles increases, TTFT grows for both methods due to more frequent cache misses and recomputation. However, TTFT grows more slowly with hybrid-aware context caching because it reuses compatible Attention KV caches and SSM state caches instead of recomputing the entire shared context. On average, hybrid-aware context caching reduces TTFT from 10.68s to 5.92s, corresponding to a 44.6\% reduction. At 20 articles, it reduces TTFT from 16.18s to 9.36s, a reduction of 42.2\%. These results demonstrate that hybrid-aware context caching improves prefill efficiency by enabling fine-grained and correct context reuse under dynamic hybrid execution.

\begin{figure}[t]
    \centering
    \includegraphics[width=\linewidth]{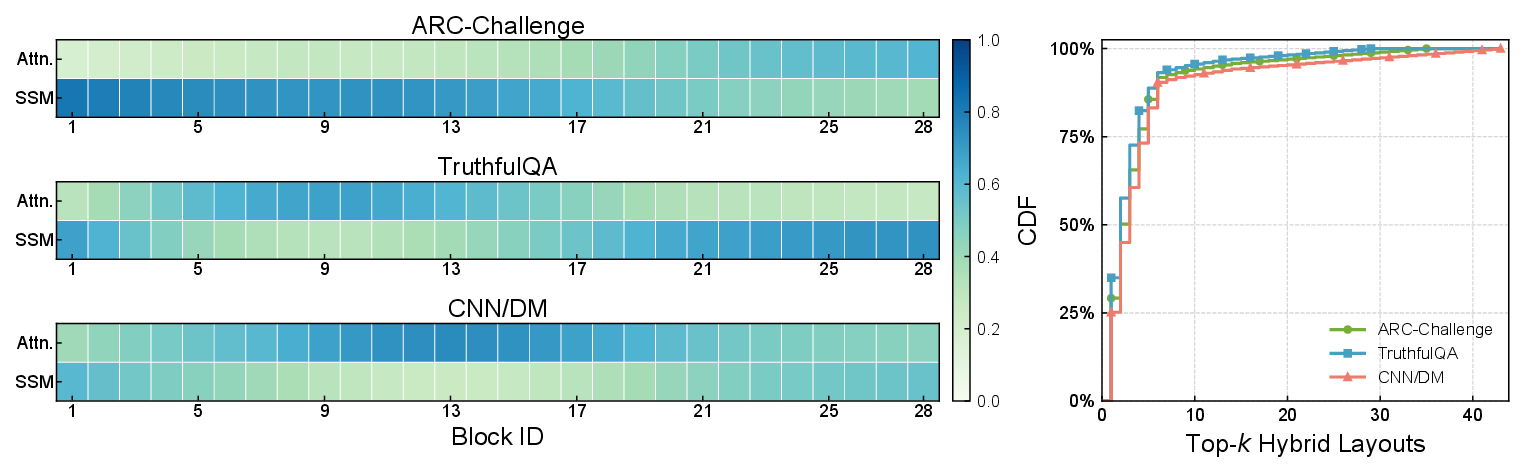}
    \caption{Routing patterns of the 28-block hybrid model across three datasets. \textbf{Left:} per-block Attention and SSM selection ratios. \textbf{Right:} cumulative request coverage by the $k$ most frequent hybrid layouts within each dataset.}
    \label{fig:abl_activation}
\end{figure}
\noindent\textbf{Layer-wise selection.} \autoref{fig:abl_activation} characterizes routing in the 28-block hybrid model in \textsc{Flex} across three datasets. The heatmaps show the fractions of requests selecting Attention and SSM at each block. On ARC-Challenge, Attention selection gradually increases toward deeper blocks. On TruthfulQA, it peaks in the early-to-middle blocks and decreases thereafter. On CNN/DailyMail, Attention selection is highest in the middle blocks and lower toward both ends. These patterns show that the allocation of Attention and SSM computation varies across both block depth and workloads.

The right panel illustrates cumulative request coverage by the top-$k$ full-model layouts, ranked by frequency within each dataset. This view complements the per-block statistics by characterizing complete routing decisions across requests. For ARC-Challenge, 500 requests activate 35 distinct layouts, with the six most frequent covering over 90\% of requests. Request-level routing therefore combines a small set of frequently selected layouts with a long tail of less common choices.

\section{Related Work}
\noindent\textbf{Hybrid LLMs.} Recent works reduce the cost of long-context LLM inference by replacing part of the attention computation with efficient sequence modeling modules~\cite{peng2023rwkv, sun2023retentive, de2024griffin}. SSM-based models, such as Mamba~\cite{gu2023mamba} and Mamba-2~\cite{dao2024transformers} maintain compact recurrent states instead of full Attention KV caches. Hybrid LLMs further combine Attention and SSM layers to balance global-context modeling and inference efficiency, with representative examples including Jamba~\cite{lenz2025jamba}, Zamba~\cite{glorioso2024zamba2}, Hymba~\cite{dong2025hymba}, Mamba-in-Llama~\cite{wang2024mamba}, and Nemotron-H~\cite{blakeman2025nemotron}. While prior hybrid LLMs demonstrate the promise of Attention and SSM hybridization, they rely on static layouts, whereas \textsc{Flex} enables request-adaptive Attention and SSM layer selection within the same model.

\noindent\textbf{Efficient batching.} Efficient batching is central to high-throughput LLM serving. Orca~\cite{yu2022orca} introduces iteration-level scheduling, while vLLM~\cite{kwon2023efficient} improves batching capacity with PagedAttention. SGLang~\cite{zheng2024sglang} supports structured LLM program execution, and systems such as Sarathi-Serve~\cite{agrawal2024taming}, Splitwise~\cite{patel2024splitwise}, DistServe~\cite{zhong2024distserve}, LoongServe~\cite{wu2024loongserve}, NanoFlow~\cite{zhu2025nanoflow}, and MuxWise~\cite{chen2026towards} optimize prefill/decode scheduling or GPU resource coordination. Unlike these systems, \textsc{Flex} supports batches with divergent Attention and SSM layer choices and preserves efficiency through selective hybrid batching.

\noindent\textbf{Context caching.} Context caching reduces redundant prefill computation by reusing caches for repeated prefixes~\cite{jin2025ragcache, yao2025cacheblend, yu2025ic, zhang2025pqcache, ye2026kvcomm}. SGLang~\cite{zheng2024sglang} proposes RadixAttention for prefix-based cache reuse, and works such as Parrot~\cite{lin2024parrot}, CacheBlend~\cite{yao2025cacheblend}, and Mooncake~\cite{qin2025mooncake} exploit reuse opportunities in structured, RAG, multi-turn, or disaggregated serving scenarios. Unlike prior techniques that bind each matched prefix to a single reusable cache, \textsc{Flex} uses the two-level radix tree to distinguish different layer-wise Attention KV and SSM state caches for the same textual prefix.

\section{Conclusion}
This paper presents \textsc{Flex}, an inference system that combines a request-adaptive hybrid model with an optimized serving runtime through algorithm-system co-design. The model uses prefill-oriented gating to select Attention or SSM branches for each request and keeps the resulting layout fixed throughout decoding. The runtime integrates selective batching and hybrid-aware context caching to support efficient mixed-branch execution and correct prefix reuse. Experiments show that the hybrid model achieves a better quality-efficiency trade-off than static hybrid baselines, while the runtime improves serving throughput and reduces inference latency.

\bibliographystyle{ACM-Reference-Format}
\bibliography{ref}

\end{document}